# General hybrid asymmetric capacitor model: validation with a commercial lithium ion capacitor

J. M. Campillo-Robles[a,*], X. Artetxe[b], K. del Teso Sánchez[b], C. Gutiérrez[c], H. Macicior[c], S. Röser[d], R. Wagner[d] and M. Winter[d,e].

[a] Fisika Aplikatua II Saila, Zientzia eta Teknologia Fakultatea, UPV/EHU, P. O. Box 644, Bilbo, Basque Country, Spain.

[b] Mekanika eta Ekoizpen Industrialeko Saila, Mondragon Unibertsitatea, Loramendi 4, 20500 Arrasate, Basque Country, Spain.

[c] CIDETEC, Pº Miramón 196, 20014 Donostia-San Sebastián, Spain.

[d] MEET Battery Research Center/Institute of Physical Chemistry, University of Münster, Corrensstrasse 46, D-48149 Münster, Germany.

[e] Helmholtz-Institute Münster, IEK-12, Forschungszentrum Jülich GmbH, Corrensstr. 46, D-48149 Münster, Germany.

*Corresponding author: Tel.: +34674062468 / FAX.: +34943791536

Email addresses: joxemi.campillo@ehu.eus (J. M. Campillo-Robles); xartetxe@mondragon.edu (X. Artetxe); kdelteso@mondragon.edu (K. del Teso Sánchez); cgutierrez@cidetec.es (C. Gutiérrez); hmacicior@cidetec.es (H. Macicior); stephan.roeser@uni-muenster.de (S. Röser); ralf.wagner@uni-muenster.de (R. Wagner); martin.winter@uni-muenster.de, m.winter@fz-juelich.de (M. Winter).

**Graphical abstract**

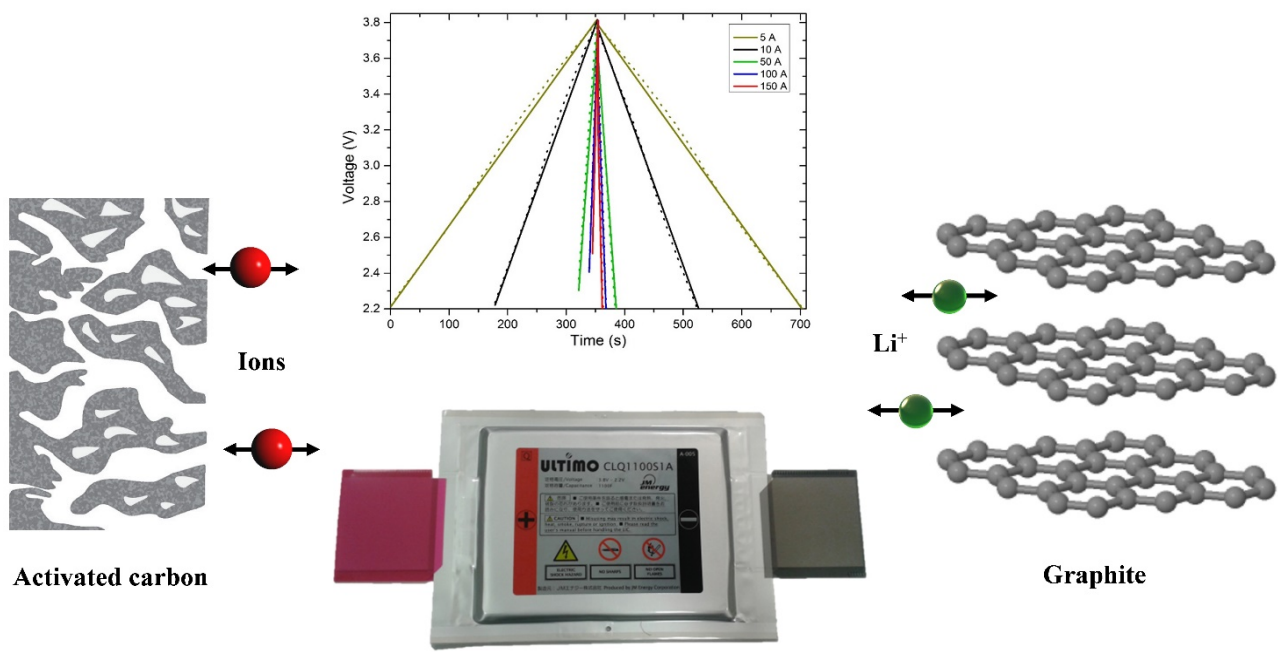

**Abstract**

Modelling and numerical simulations play a vital role in the design and optimization of electrochemical energy storage devices. In this study, a general physics-based model is developed to describe Hybrid Asymmetric Capacitors (HACs). A one-dimensional cell is constructed with one faradaic electrode, a separator and a capacitive electrode. The model is validated using a commercial Lithium Ion Capacitor (LIC). Galvanostatic charge and discharge processes are simulated with a maximum mean relative error of 7.8%. This suggest that this simple Ohmic model captures the key electrochemical phenomena occurring inside the LIC cell.

Research Highlights:

- A general physics-based model is developed in 1D to simulate HAC.

- A commercial LIC is used to validate the general ohmic model.

- Numerical results reproduce accurately experimental electrical features of the LIC.

- Maximum mean relative error achieved in galvanostatic processes is less than 7.8%.

# 1. Introduction

The depletion of fossil fuels and environmental issues made renewable energy generation a necessity for a sustainable development of our society [1]. Furthermore, hybrid and electric transportation is increasing throughout the world [2]. Due to this, important progresses of the electrical grid, such as grid scale energy storage or smart grid, are under ongoing development [3]. This global energy transition leads to an increasing need of new and more effective as well as reliable energy storage devices [4]. This need is further related to the rapid technological advances of the last decades, in the field of electronics, robotics, informatics and communication. For all these reasons, researchers are continuously developing new promising electrochemical energy storage technologies, such as, sodium ion batteries, metal-air batteries, flow batteries, etc. [5]. In this context, hybridization has become an interesting attempt to improve the energy and power densities of the energy storage devices. Hybrid asymmetric capacitor (HAC) is an example of this kind of novel devices [6-8].

A HAC combines a capacitive electrode usually used in electrochemical double layer capacitors (EDLCs) and a battery-type faradaic electrode with redox reaction or insertion. There are two major approaches to select the faradaic electrode of a HAC [9]: a) pseudocapacitive metal oxides ($RuO_2$, $Ni(OH)_2$, $MnO_2$, $V_2O_5$ [10], etc.) or b) lithium-insertion electrodes ($LiMn_2O_4$, graphite, $Li_4Ti_5O_{12}$, etc.) [11]. In 1991, Conway developed the first pseudocapacitive HAC using hydrated ruthenium oxide as faradaic electrode [12]. Twenty years later, in 2001, Amatucci *et al.* reported, named and patented the first HAC with lithium insertion electrodes [13, 14]. They used a nanostructured $Li_4Ti_5O_{12}$ negative electrode combined with an activated carbon (AC) positive electrode, both immersed in an organic electrolyte [13]. In recent years, many chemistries and configurations have been reported for the application in these hybrid devices [15-21].

## 1.1. Lithium ion capacitors

Today, Lithium-Ion Capacitors (LICs) are the most competitive devices among HACs [22, 23]. The principal characteristic of these devices is that they have a lithium-ion insertion electrode. LICs have some advantages, in contrast with other rechargeable battery chemistries: long cycle lifetime expectance, reliability and superior energy efficiency. In fact, the energy and power delivery performance of LICs fills the gap between EDLCs´ low energy density and traditional batteries´ low power density. For this reason, the leading applications of LICs are increasing day by day [24]. LIC applications can be principally divided into four groups: storage, energy regeneration, leveling and backup systems [25]. Nowadays, several manufacturers have started their commercialization: JM Energy, Yunasko, General Capacitor, Taiyo Yuden, etc.

In a LIC, the negative faradaic electrode usually consists of a lithium-ion insertion electrode typical of Lithium-Ion Batteries (LIB): graphite (pristine), $Li_4Ti_5O_{12}$ or hard carbon. Actually, prelithiated graphite is particularly attractive due to its low negative redox potential (0.1 V versus Li/Li$^+$) and high theoretical capacity (372 mAh g$^{-1}$) [26]. A LIC cell with graphite as a negative electrode and the commonly used AC of EDLCs as a positive electrode will have greater energy storage capacity than typical EDLCs, because of the higher capacity of graphite, combined with an increased operating voltage (2.2-3.8 V). As a result, most of the commercially available devices use this configuration.

Figure 1 shows the schematic evolution of potentials in a LIC cell during cycling of the device. As mentioned previously, the potential of a graphite electrode in a LIC remains almost constant at around 0.1 V versus Li/Li$^+$ during cycling of the device (due to the preceding prelithiation process and the oversize of this electrode with respect to positive electrode) [27-29]. AC electrode has a working potential window of 1.6-4.6 V versus Li/Li$^+$ [27]. However, to obtain long cycle life, in commercial LICs the potential of positive electrode is only allowed to vary between 2.3-3.9 V versus Li/Li$^+$ [30]. Therefore, the working voltage window for the device is 2.2-3.8 V.

As like in LIBs [31, 32], the electrolyte of commercial LIC consist mainly of LiPF$_6$ salt dissolved in a mixture of organic solvents. During operation, anions and cations of the electrolyte undergo different processes [33]. At the negative electrode, insertion and extraction of lithium ions take place. Whereas, at the positive electrode, absorption and desorption of ions (Li$^+$ and PF$_6^-$) happens. More specifically, during the initial stage of the discharge process (3.8 V to 3.0 V) Li$^+$ ions are extracted from the graphite electrode and PF$_6^-$ ions are desorbed from AC surface. For this reason, the concentration of $Li^+$ and $PF_6^-$ ions in the electrolyte increases, and, as a consequence. Furthermore, during the 3.0 V to 2.2 V discharging process, the $PF_6^-$ ion does not take part in the charge-transfer process (its average concentration remains constant in the electrolyte). Meanwhile, $Li^+$ ions are extracted from the graphite electrode, and are absorbed at the AC surface. Consequently, the average concentration of $Li^+$ and $PF_6^-$ ions remains constant in the electrolyte in the voltage range of 3.0 V to 2.2 V. Inverse processes proceed during the charge operation at the given voltage ranges. Figure 1 summarizes all these charge-transfer processes. The concentration changes are more important near the electrode-electrolyte interfaces, especially, if the applied current is high.

## 1.2. Modelling of capacitors

Future advances in the development of new energy storage devices will be greatly influenced by the need to engineer materials and processes using advanced simulation tools. In the last decades, a huge effort has been performed to develop models for different electrochemical systems [34]. Newman and Tiedemann were among the pioneers of exploiting numerical techniques for solving the governing equations of both primary and secondary batteries in the 1970s [35, 36]. Afterwards, lead-acid batteries [37] and LIBs [38-40] have been widely modeled and simulated due to their commercial importance. Likewise, different models for EDLCs have also been accomplished. From the first models of Helmholtz [41], Gouy [42], Chapman [43] and Stern [44], significant progress has been made modeling EDLCs [40, 45-47]. Moreover, the mathematical framework of EDLC models can be extended to explain HACs with electrodes featuring redox reactions, with or without ion intercalation. In 2014, Staser and Weidner developed a one-dimensional model for an asymmetric capacitor of nickel chemistry in aqueous electrolyte. Later, they expanded this model to lithium chemistry in an organic electrolyte [48]. In this model, for all cells AC was used as the cathode and a non-porous metallic foil (Ni or Li) for the faradic electrode. In the following, Hao *et al.* presented a model to describe a $LiMn_2O_4$/AC capacitor with 1 M $Li_2SO_4$ aqueous electrolyte. The validated model was used to optimize the size of the cell size in various applications [49]. Subsequently, d´Entremont *et al.* performed numerical simulations of a HAC with planar electrodes ($Li_2Nb_2O_5$-carbon) and 1 M $LiClO_4$ organic solvent electrolyte [50-52]. In the case of LICs, all the modeling work has been devoted to develop RC equivalent-circuit models [53-55], which however have limited prediction capacity. To address these drawbacks, we recently have presented preliminary results of a physics-based model for LICs [56]. The purpose of the present work is to introduce an electrochemical model valid for all kind of HACs.

Firstly, a description of the general model developed to explain the electric behavior of HACs is presented. Secondly, we present the technical characteristics of the selected commercial LIC we are going to use for validation of the model and the parameters used in the simulation. Then, the numerical procedure is explained. Afterwards, numerical results and measured galvanostatic charge-discharge processes are compared to check the accuracy of the model describing the electrical behavior of the LIC. We finalize our research with the discussion of the assumptions and limitations of the model.

## 2. General model description

The one-dimensional electrochemical model used to describe LICs derived in Ref. [56] is an ohmic porous-electrode model. However, it can also be generalized to describe all kind of capacitors, from HACs to EDLCs. Its application to any capacitor requires the characteristics of the materials of both electrodes, and the electrolyte. The general model can be summarized as follows.

### 2.1. Model assumptions

We have adopted some simplifying assumptions to develop the general electrochemical model:

1) The model is one-dimensional; i.e. we have supposed that current distribution along the height or the width of the electrodes remains uniform.
2) All the properties of the materials are uniform and constant during cycling of the cell. Therefore, conductivities, porosities, active material capacitance and electrolyte concentration are uniform, and they remain constant over the voltage operation window.
3) The temperature is uniform and constant in the cell, and equal to 298.15 K (isothermal model approach).

4) The energy storage mechanism is completely arising from the absorption/desorption of ions in the capacitive electrode, and due to redox reaction or intercalation in the faradaic electrode (in LICs, $Li^+$ insertion/extraction).

5) The current in the solid-phase is electronic, while in the liquid-phase is fully ionic (see Figure 2). In both phases, charge transport has been modeled using Ohm´s law. No mass transport has been modeled in the system (no diffusion).

6) No potential drop appears in current collectors (perfect conductors). This means that the ends of the active materials act as boundaries of the system. Other non-electrode components of the cell are considered to have negligible resistance, and they do not contribute to the changes in the voltage of the cell.

## 2.2. Geometry

The generic cell is described with a single cell sandwich level approach; i. e., the model has three domains with two porous electrodes, separated by an ionically conductive but electronically insulating separator. Figure 2 shows a schematic diagram of the cell. The cell consists of the following domains and boundaries:

(a) boundary: current collector of the negative electrode ($x = 0$);

(b) domain: negative faradaic electrode ($0 < x < L_-$);

(c) boundary: negative electrode-separator ($x = L_-$);

(d) domain: separator ($L_- < x < L_- + L_{sep}$);

(e) boundary: separator-positive electrode ($x = L_- + L_{sep}$);

(f) domain: positive capacitive electrode ($L_- + L_{sep} < x < L_- + L_{sep} + L_+$);

and (g) boundary: current collector of the positive electrode ($x = L_- + L_{sep} + L_+$).

This general model has been applied to a commercial LIC cell, which has a negative electrode of graphite, an electrically insulating polymeric separator and a positive electrode of AC. These three domains are porous; with different porosity values related to each material (see Section 3). These three porous materials are impregnated with non-aqueous electrolyte of a lithium salt.

## 2.3. Governing equations

In the following lines, the governing equations for developing the 1D general electrochemical model are presented. The volume-averaged porous electrode approach has been used in electrochemical modeling for years [57, 58]. The two electrodes, faradaic and capacitive, are described with the well-known porous electrode theory, originally derived by Newman and Tiedemann in 1975 [35, 36].

There are two physical dependent variables in the model: the solid phase potential, $\phi_s$, and the electrolyte phase potential, $\phi_e$. Electrolyte phase potential, $\phi_e$, is defined in the three domains, whereas solid phase variable, $\phi_s$, is only used in the two electrodes domains. The independent variables of the model are the spatial coordinate $x$ and time $t$.

The conservation of charge in the cell is described as the sum of the current density crossing through the ionic conductive phase (electrolyte, liquid phase), $j_e$, and the electronically conductive phase (electrode matrix, solid phase), $j_s$:

$$j_{cell} = \frac{I}{A} = j_e + j_s .$$ (1)

where $j_{cell}$ is the total current density of the cell (A cm$^{-2}$). This total current density is the intensity that crosses the cell, $I$ (A), divided by the cross sectional area of the cell, $A$ (cm$^2$). Ohm's law relates current density to the gradient of the electric potential in the solid and liquid phases:

$$j_e = -\kappa \frac{\partial \phi_e}{\partial x} ,$$ (2)

$$j_s = -\sigma \frac{\partial \phi_s}{\partial x} ,$$ (3)

where $\kappa$ and $\sigma$ are the effective conductivity of the electrolyte and solid phase, respectively (S m$^{-1}$).

In order to calculate the effective conductivity of the electrolyte and the solid matrix, we have used the well-known Bruggeman relation [59, 60]. This equation gives a relation for the effective conductivity in function of the bulk conductivity of the electrolyte, $\kappa_0$ (solid phase, $\sigma_0$) and the volume-fraction of the electrolyte, $\varepsilon_e$ (solid phase, $\varepsilon_s$). We have taken the frequently used value of 1.5 for the Bruggeman´s exponent [57, 58, 61].

$$\kappa = \kappa_0 \varepsilon_e^{1.5} \,,$$
(4)

$$\sigma = \sigma_0 \varepsilon_s^{1.5} \,.$$
(5)

Finally, taking into account Equations (1) to (3), the conservation of charge can be summarized as:

$$\frac{\partial j_s}{\partial x} = -\frac{\partial j_e}{\partial x} = a j_D \,,$$
(6)

where $a$ is the specific electroactive area of the electrodes (cm$^2$ cm$^{-3}$) and $j_D$ is the pore wall flux (A cm$^{-2}$). Thus, $a j_D$ represents the volumetric current density between the solid matrix and the liquid phase in the electrode (A cm$^{-3}$).

### 2.3.1. Capacitive electrode

The capacitive electrode has been modeled assuming that the only energy storage mechanism is the double layer effect. Therefore, the pore wall flux between the porous electrode matrix and the electrolyte is described as:

$$j_D = -C\left(\frac{\partial(\phi_s - \phi_e)}{\partial t}\right) = -C\frac{\partial \eta_+}{\partial t} \,,$$
(7)

where $j_D$ is the total double-layer current density (A m$^{-2}$); $C$ is the electrical double layer capacitance of the active material of cathode (F m$^{-2}$). The $\eta_+$ overpotential (V) of the capacitive electrode is defined as:

$$\eta_+ = \phi_s - \phi_e \,.$$
(8)

### 2.3.2. Separator

An electronically insulating porous separator is necessary to prevent short circuits between oppositely polarized electrodes in the capacitor. The depletion of the electrolyte conductivity due to the separator has been modeled using the Bruggeman relation with a coefficient of 1.5 [62]:

$$\kappa = \kappa_0 \varepsilon_{sep}^{1.5},\tag{9}$$

where $\varepsilon_{sep}$ is the volume fraction of the electrolyte in the separator domain.

### 2.3.3. Faradaic electrode

At the faradaic electrode of a HAC, redox reaction or intercalation process can happen. In the theoretical case of LICs, the only mechanism of energy storage is the Li$^+$ insertion/extraction process. The well-known Butler-Volmer equation has been used to describe the charge-transfer reaction between the active material and electrolyte. The pore wall flux is described as:

$$j_D = j_0 \left[ \exp\left( \frac{\alpha_a F \eta_-}{RT} \right) - \exp\left( \frac{-\alpha_c F \eta_-}{RT} \right) \right],\tag{10}$$

where $j_0$ is the exchange current density (A m$^{-2}$); $\alpha_a$ and $\alpha_c$ are the anodic and cathodic charge transfer coefficients; $F$ is the Faraday constant: 96,485 C mol$^{-1}$; $R$ the universal gas constant, 8.3143 J mol$^{-1}$ K$^{-1}$ and $T$ the temperature in K.

The exchange current density depends on both, lithium concentration at the surface of the electrode and the lithium concentration of Li$^+$ ions in the electrolyte adjacent to the interface. However, in this model, we have set the exchange current density as a constant based on the experimental data available in the bibliography [63, 64]. This means, that changes in the lithium concentration in liquid and solid phases have not been taken into account.

Moreover, the anodic and cathodic charge transfer coefficients fulfill the relation $\alpha_a + \alpha_c = 1$. These coefficients are weighting the anodic and the cathodic contribution of the current to the overall reaction induced by the overpotential $\eta_-$ (V):

$$\eta_- = \phi_s - \phi_e - U_{eq} - \Delta\phi_{film}.\tag{11}$$

In this expression, $U_{eq}$ is the equilibrium potential of the faradaic electrode (V). Moreover, $\Delta\phi_{film}$ is the ohmic drop (V) due to the film resistance of the faradaic electrode. This ohmic drop is expressed as:

$$\Delta\phi_{film} = R_{film} j_D \,,\tag{12}$$

where $R_{film}$ ($\Omega$ m$^2$) is mainly related to the ionic conductivity of the solid electrolyte interphase (SEI) layer formed initially, during the prelithiation process of the negative electrode [65, 66].

## 2.4. Boundary and initial conditions

The simulated cell is a closed system. Therefore, there is no mass flow through the walls. On the boundaries, all the electric current is in the solid phase (see Figure 2); consequently, the electrolyte potential fulfills the condition:

$$\left.\frac{\partial\phi_e}{\partial x}\right|_{x=0,\, L_- + L_{sep} + L_+} = 0 \tag{13}$$

The boundary conditions for the electric potential of the solid phase at the current collectors are the next ones (see Figure 1):

$$\begin{cases} \phi_s = 0 & \text{Fixed potential at current collector of negative electrode } (x = 0) \\ -\sigma\dfrac{\partial\phi_s}{\partial x} = \dfrac{I}{A} & \text{Fixed current at current collector of positive electrode} \end{cases} \tag{14}$$

Negative values of $I$ refer to charging of the cell, whereas positive values of $I$ denote discharging.

For solving the governing equations of the cell, initial values for the two dependent variables are necessary. Initially, the electrolyte potential was taken as uniform and equal to zero across the entire cell, $\phi_e = 0$. The solid potential at the current collector of the positive electrode, $\phi_s(L_- + L_{sep} + L_+)$, has two different initial values, depending on the initial state of charge (SOC) of the cell: charged (3.9 V, SOC = 100%) or discharged (2.3 V, SOC = 0%). For the rest of the cell, the solid potential was taken as uniform and equal to zero $\phi_s = 0$.

## 3. Parameters of the model

In this work, the electrical tests have been performed on a commercial available LIC. It is a pouch-type cell manufactured by JM Energy: Ultimo 1.100 F [67]. The relevant parameters of the commercial LIC are provided in Table 1. Each LIC cell has 17 double-sided electrode-pairs [68]. The coatings of the capacitive and faradaic electrodes and the separator thickness are listed in Table 2. Moreover, the dimensions of the electrodes of this pouch cell give an overall cross sectional area of 0.4352 m$^2$. The rest of the parameters used in the simulations are depicted in Table 2. Most of them have been collected from experimental measurements performed in the literature.

The electrolyte of the commercial LIC pouch cell is 1 M LiPF$_6$ in ethylene carbonate : ethyl methyl carbonate : dimethyl carbonate (EC:EMC:DMC = 1:1:1, by weight) solvent. A sample of this electrolyte has been prepared to measure the conductivity. Conductivity values of the electrolyte were determined by impedance measurements using a potentiostat/galvanostat PGSTAT302N (Methrom Autolab). The investigated temperature range was from -30 to 60 °C with $\Delta T$ = 10 °C. The temperature was controlled to ±0.1 °C by a Microcell HC temperature interface from rhd instruments. The voltage amplitude was set to 10 mV in a frequency range of 100 – 1 kHz. A total of 1 ml of electrolyte was placed into a measuring cell (TSC 1600 closed, rhd instruments) using glassy carbon ($d$ = 3 mm) and platinum as electrode setup inside a glovebox (O$_2$ and H$_2$O contents < 0.5 ppm). Both electrodes were polished, rinsed clean and dried prior each measurement. The cell constant was determined using 0.01 M KCl reference solution (VWR) with a defined conductivity of 1.413 µS cm$^{-1}$ at 25°C. Repetition of the experiments gave an uncertainty of ±0.1 mS cm$^{-1}$ of the obtained results.

We have used MATLAB 2017a[®] [78] for fitting a polynomial equation to the conductivity data using a least-squares linear regression method. We have chosen the maximum degree of the polynomial equation, analyzing the coefficient of determination, $R^2$. A second-degree polynomial gives a value for $R^2$ of 0.9986. Increasing the polynomial degree to three, $R^2$ gets a value of 1. Therefore, we have chosen a third degree polynomial to describe the conductivity of the electrolyte, similarly to that presented by Ding et al. [79]. Data has been normalized before the fitting. With data normalization, we avoid an ill-conditioned matrix in the fitting of Equation (15). Moreover, the normalization does not affect the error of the fitting or the value of $R^2$. Consequently, we have obtained the following equation for the conductivity in the temperature range of -30-60 °C:

$$\kappa = 206.19 - 2.3476\,T + 8.5614 \times 10^{-3}\,T^2 - 9.6813 \times 10^{-6}\,T^3 , \tag{15}$$

where $T$ is expressed in K and $\kappa$ in mS cm$^{-1}$. Figure 3 shows the ten measured values of the conductivity and the fitted Equation (15).

## 4. Numerical simulations

The governing equations of the 1D model and the associated initial and boundary conditions have been numerically solved using finite element methods. We have used the secondary current distribution interface of the commercial finite-element solver COMSOL Multiphysics® (version 5.3) [80]. The two dependent variables, $\phi_s$ and $\phi_e$, have been discretized using quintic order elements. Multifrontal Massively Parallel Sparse Direct Solver (MUMPS) has been chosen as linear solver, with Backward Differentiation Formula (BDF) time stepping method. The absolute tolerance for convergence has been below $10^{-3}$ for all variables. In order to ensure the accuracy of the solution, we have used a non-uniform 1D mesh, increasing the mesh density near each domain´s boundary (maximum element size: 1.05 μm). For low rate charge-discharges (5-80 A) the time-step has been fixed at 0.1 s, and for fast charge-discharge processes (100-350 A) at 0.01 s. In all cases, we have checked that the numerical results are independent of mesh density and time-step.

## 5. Results and discussion

The results of the present research work are explained in this section.

### 5.1. Experimental measurements

In order to test the suitability of the model, the commercial pouch LIC cell (Ultimo 1.100 F) has been tested using a commercial battery cycler (Battery tester Basytec model HPS [81]) at constant temperature, 25 °C (temperature chamber Vötsch model VTS 4034-5). We have checked all the working range of intensities of the LIC: charging (5, 10, 20, 30, 40 and 50 A) and discharging (5, 10, 30, 50, 80, 100, 150, 200, 250, 300 and 350 A). The limited rate performance of conventional LICs

is due to the graphite anode [82]. Indeed, the charging process is limited in LICs, because current densities near or above 4 mA/cm$^2$ on charge should be avoided in graphite cells, unless additional precautions have been taken to avoid side reactions [83]. For this reason, the charging rate of this commercial pouch cell is limited to a maximum of 50 A [67].

A selection of the measured galvanostatic discharge profiles are plotted in Figure 4 (voltage vs. time). It can be observed that the cell voltage, likewise to an EDLC, shows a nearly linear behavior. As a result, LIC voltage has a nearly linear relationship with the SOC. In this case, cell voltages of 3.8 V, 3.5 V, 2.9 V, 2.6 V and 2.3 V approximate a SOC of: 100%, 70%, 45%, 25% and 5% [54]. Table 3 collects the most notable electrical properties of the measured galvanostatic discharging processes.

Nevertheless, discharge curves are not perfectly linear. All of them show a little slope change near 3.0 V; below this voltage the curve slope is slightly higher. For this reason, we have compared the measured voltage with a hypothetic linear behavior. To perform this comparison, we have made a linear fit of the measured voltage data for all the discharging processes (worst fit $R^2$ = 0.9954). In continuation, we have calculated the relative deviation between the measured voltage and the linear fit (see Figure 5). The deviation shows a nearly sinusoidal behavior with a maximum of +1.0% and a minimum of -2.5%. The maximum of the voltage deviation shifts to the right (longer time) as the discharge intensity increases (see Figure 5). This could be explained due to the slower kinetic of Li$^+$ insertion/extraction process with respect to the double layer mechanism. This effect is more evident at high C-rates. At 3.8 V, the experimental voltage value is always smaller than the linear fitted value, due to the fact that the voltage response at elevated voltages has a smaller slope. The electrolyte conductivity inside the LIC is the lowest one at 3.8 V for all the voltage range (lowest concentration of the conductive salt). As the concentration of the Li$^+$ and PF$_6^-$ ions in the electrolyte rises from 3.8 V to 3.0 V (increase in conductivity), the experimental voltage value deviation increases, and exceeds the linear value. During the 3.0 V to 2.2 V discharging process, the electrolyte concentration remains constant at a maximum, and the charge transfer happens only because of the Li$^+$ movement. As a result, the experimental voltage deviation decreases to a minimum, ceasing anew with a slight increase.

Finally, representative examples of the galvanostatic charging processes also have been plotted in Figure 6 (with a maximum intensity of charging processes limited to 50 A). As expected, the voltage of the LIC device practically shows linear behavior for all the intensities.

## 5.2. Validation of the model

The simulated voltage of discharging processes of the LIC has been plotted in Figure 4. As mentioned before, this simplified model uses important simplifying assumptions: constant electrolyte concentration, negligible ohmic resistance of other elements of the cell, only ohmic transport (no diffusion) and constant potential of the lithiated graphite electrode among others. For this reason, this ohmic model is not appropriate to study the inner dynamics of the device (i.e. concentration changes, ion diffusion-migration processes, overpotential evolution, insertion-extraction of $Li^+$).

Due to these simplifications, the simulated discharge curves presents a linear behavior (see Figure 4). Considering the simplicity of the model, the correlation between the experimentally obtained and numerically simulated data is high. Table 3 shows that the maximum and mean relative deviation in the voltage increases as the applied current increases. The maximum mean relative deviation of the voltage appears at 350 A discharge (7.8%). It can be seen from Figure 4 that the maximum deviation appears at low C-rates. Nevertheless, at high C-rates the slope of the discharging profile is very high, and, as a result, the vertical difference between the experimental and the simulated profiles gets higher. Indeed, the horizontal deviation between the curves at high C-rates (deviation of time for the same voltage) is smaller than the deviation of the voltage. Furthermore, initial voltage of the discharge is described quite well for low rate discharges (see Table 3), but at high rates the simulated values increase more than experimental ones. The highest discrepancy is 4.2% at 350 A.

Figure 6 shows the measured and numerically simulated charge curves. The linear behavior of charging processes show a very high correlation with the experimental curves. The relative deviation of the voltage is similar to that of the discharging processes. Due to the current limitation (50 A), the maximum relative error of the voltage is not greater than 3.5%.

## 5.3. Peukert´s law

In 1897, W. Peukert developed a simple relation to predict the amount of energy that can be extracted from a cell [84]. This widely used empirical formula describes the change in the capacity of a cell at different current rates of discharge as expressed in Equation (16):

$$C_{1A} = I^p t, \tag{16}$$

where $I^p$ represents the galvanostatic discharge current (A), $t$ is the discharge time (h), $C_{1A}$ is the theoretical capacity transferred by the cell at 1 A current (Ah) and $p$ an empirical dimensionless parameter called Peukert´s coefficient. A $p$ value close to unity represents an efficient cell (the discharge capacity is independent of the applied current), and it will increase as the cell efficiency decreases. Moreover, $p$ increases with decreasing temperature of the device. This fact shows the capacity/efficiency reduction of the cell at low temperatures. Originally, Peukert´s law was developed for lead acid batteries [84, 85], but its validity has expanded to other electrochemical storage devices, such as LIBs [86], NiMH batteries [87] and LICs [88-90]. Moreover, recently, the validity of Peukert´s law has been proven in EDLCs too [91, 92]. Therefore, Peukert´s law can be used to predict discharging times at different constant intensities for most of the commonly used energy storage devices. Unfortunately, the accuracy of Peukert´s law lessens under dynamic loading and varying temperature conditions. For this reason, some approaches have been performed to improve the prediction capacity of Peukert´s law, for a wider range of intensity and temperature [93-95]. However, the resulting relation is more difficult for practical applications.

Experimentally obtained discharge times and intensities of the commercial LIC have been plotted on a logarithmic scale (see Figure 7). An obvious reduction of the LIC capacity increasing the current can be observed. Based on the results of Table 3, we have plotted in Figure 7 the Peukert´s relation for all the experimental discharges (5-350 A) of the LIC. We have obtained a value for the Peukert´s constant of 1.084 ($R^2$ = 0.9988). Nevertheless, we also observed that $p$ value decreases considerably if we remove high C-rates discharges. Reducing the range of the currents to 5-150 A, Peukert´s constant value decreases up to 1.050 ($R^2$ = 0.9997). This shows that Peukert´s coefficient is highly dependent on the applied current maximum. Indeed, Peukert´s law is not applicable at high C-rates, because the behavior of the cell does not follow a logarithmic relation, due to the electrochemistry of the cell [95]. Nonetheless, our experimental results are in line with those of the literature. Fleurbaey et al. have measured Peukert´s constant in a 1500 F LIC from -10 ºC to 60 ºC. They found a value of 1.05 at 25 ºC using intensities from 10 A to 180 A [89]. Omar et al. quantified Peukert´s constant to 1.03 using the range of intensities of 20-200 A at 25 ºC for a 3300 F LIC [90].

In addition, Figure 7 presents Peukert´s relation for the simulated discharge times. Theoretical values are in good accordance with the experimental ones. However, at high C-rate discharges a small deviation is realized. This can be explained with the fact that simulated discharge times are longer than experimental ones (see Table 3).

## 5.4. Capacitance

There are many ways to measure the capacitance of a capacitor: constant current discharge method, electrochemical impedance spectroscopy (EIS), chronoamperometry, cyclic voltammetry (CV), among others [96, 97]. The resulting value for the capacitance is different depending on the method and the testing conditions. In fact, the capacitance depends on the discharge current and the voltage window.

In constant current discharge method, we can calculate the capacitance from Equation 17:

$$C = \frac{I \Delta t}{\left( V_0 - V_{end} \right)},$$

(17)

where $C$ is the capacitance of the device (F), $I$ the discharge current (A), $\Delta t$ discharge time (s), $V_{end}$ the rated lower limit voltage (V) and $V_0$ the instant drop voltage at discharge (V), calculated using a least-squares regression line. Equation (17) and similar have been used also in LICs [55, 88].

Table 3 shows that the capacitance of the commercial device reduces as the discharge current increases. This is a normal behavior of EDLCs [92, 95] and LICs [55, 88, 95, 98], due to the voltage drop caused by internal resistance of the cell, the reduced recovery rate and the reduced amount of ions contributing to the charge transference at high current densities. The capacitance obtained in simulations does not show this behavior, and it remains nearly constant for all the discharging intensities (see Table 3). This is related to the constant electrical double layer capacitance of the capacitive electrode we have used in the simulations.

## 5.5. Temperature effect

In comparison to LIBs, LICs have a wider working temperature range (0 °C / 50 °C, -30 °C / 70 °C, respectively). Nevertheless, the impedance of LICs is severely affected at temperatures below 10 °C [54]. It is well known that temperature affects the electrolyte conductivity (see Figure 3). At high temperatures, the movement of lithium ions and therefore charge is faster, and, as a result, the kinetic is higher. Moreover, temperature also appears in the Butler-Volmer equation applied in this model.

Although the model is isothermic, we have performed some simulations varying the temperature. At low discharging rates, the simulated voltage profile of the LIC cell is not affected noticeably by the temperature. For instance, Figure 8 a)) shows the first 4 s of a 5 A discharging process for different temperatures. The differences are very small. Nevertheless, at high discharge rates, the temperature effect is more noticeable. For example, in Figure 8 b) the voltage profile of a 350 A discharge is plotted for different temperatures. At 20 °C, 40 °C and 60 °C, the voltage profiles are nearly linear and very close to each other. However, reducing cell temperature, the electrolyte conductivity decreases (see Figure 3) and with that, the discharge capacity clearly decreases. Moreover, the voltage of the LIC does not show a linear correlation.

## 6. Conclusions

- The equations of this general model are independent of device specifications, and are valid for porous electrodes in which the main storage mechanism in the anode is a redox reaction or insertion/extraction process, whereas the main process of the cathode is the lithium absorption/desorption.

- During cycling of the cell, the properties of the materials change in space and time. The greatest deviations appear in the electrolyte concentration (and the properties related to it; i. e., electrolyte conductivity,…) and the properties affected by the cationic insertion in the negative electrode. Therefore, the constant value assumption for the properties of the materials is not useful to explain the behavior of the materials during cycling. Nevertheless, it gives a good electrical description of the entire cell.

- This 1D model is practical to predict the electrical behavior of LICs, whereas it is not adequate to analyze the transport phenomena of the cell.

- Our simulations reproduce correctly the experimental obtained data using a commercial LIC. This suggests that the ohmic model captures the key physical phenomena occurring in the device.

## 7. Acknowledgements

L. Oca (MU) is gratefully acknowledged for the work performed in the preliminary phase of this research. J. Ajuria (CIC Energigune) is thanked for kindly help in the inner characterization of the commercial LIC cell. Authors also wish to thank the support received from Mondragon Unibertsitatea (MU), Mechanical and Industrial Manufacturing Department.

**Figure captions**

**Figure 1.** Schematic depiction of potential profiles of negative and positive electrodes in a charging-discharging process of an asymmetric supercapacitor based on AC/Li-doped graphite.

**Figure 2.** Schematic view of the simulated capacitor cell (1D) and the associated coordinate system, with the current distribution throughout the cell: solid (red), liquid (blue) and exchange current densities.

**Figure 3.** Temperature dependent conductivity of 1 M solution of $LiPF_6$ in 1:1:1 EC-DMC-EMC. Conductivity values have been fitted with Equation (15).

**Figure 4.** Experimental and simulated galvanostatic discharge curves of the LIC.

**Figure 5.** Relative deviation between experimental values of voltage and the fitted linear function of the experimental data in galvanostatic discharging processes.

**Figure 6.** Experimental and simulated galvanostatic charging curves of the LIC.

**Figure 7.** Peukert´s law of the discharging processes: experimental (black circles) and simulation (red diamond). Plotted line indicates the least squares fit of all the experimental values.

**Figure 8.** Temperature effect on the discharging processes: a) 5 A (first 4 s) and b) 300 A.

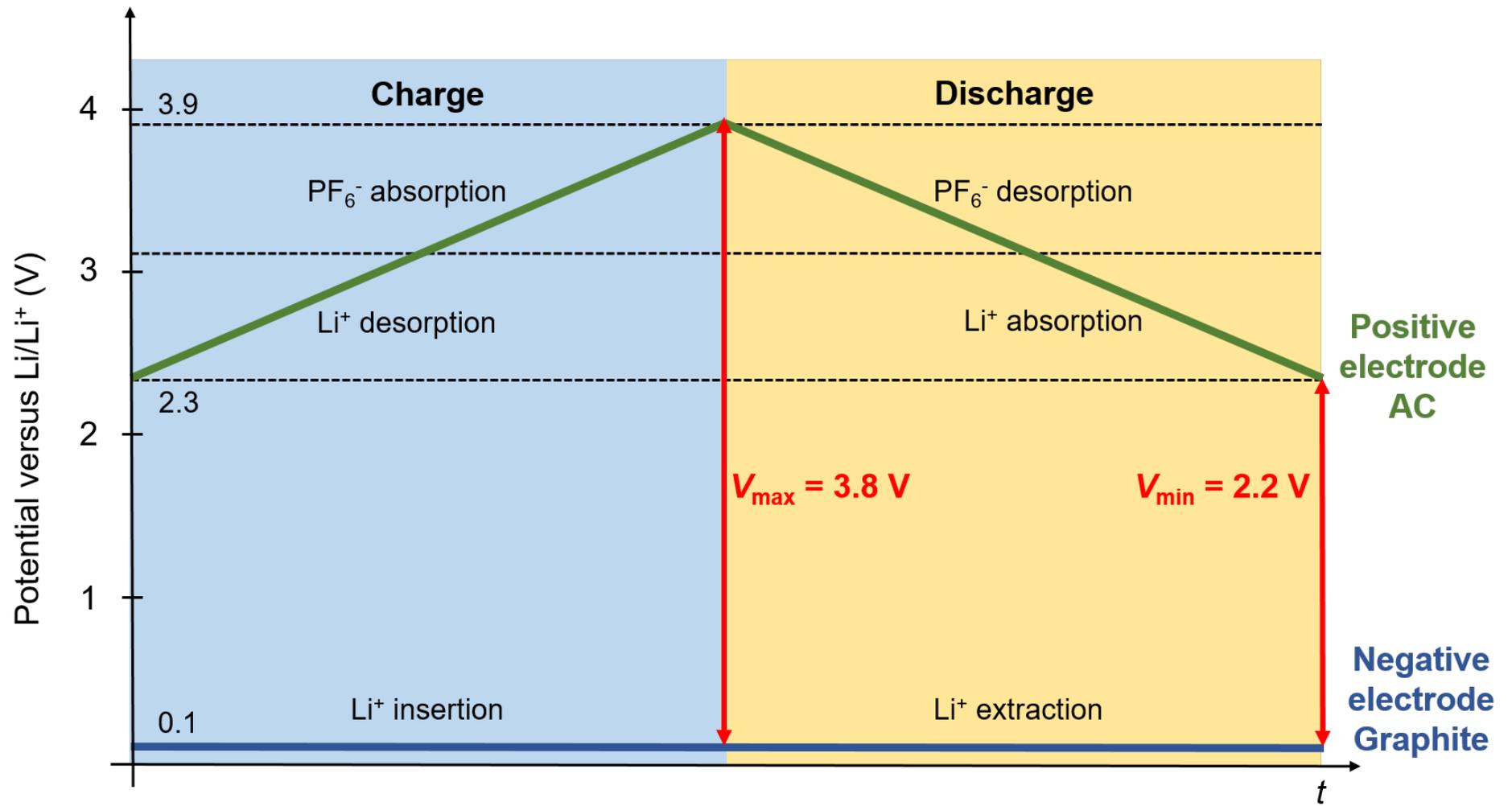

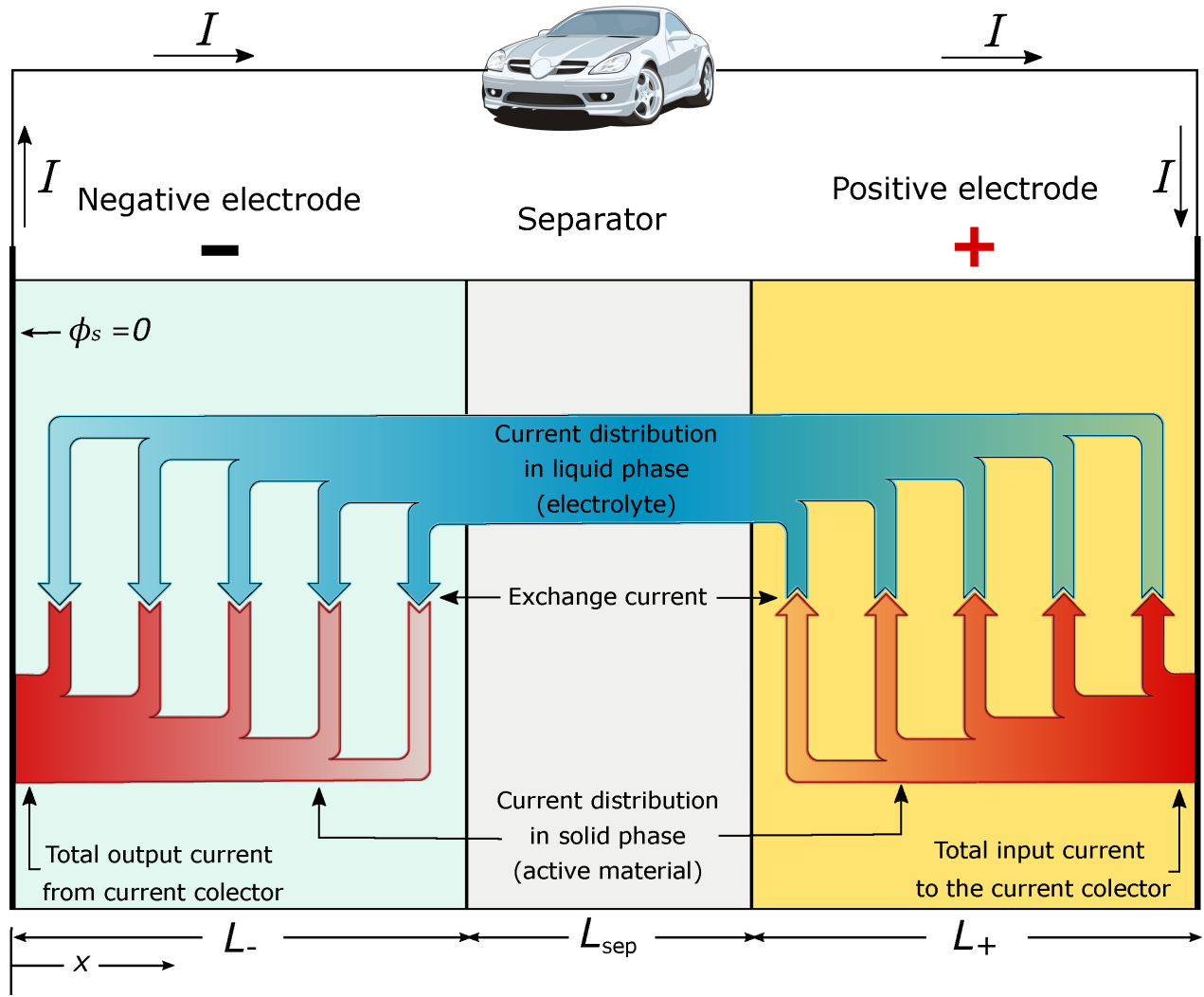

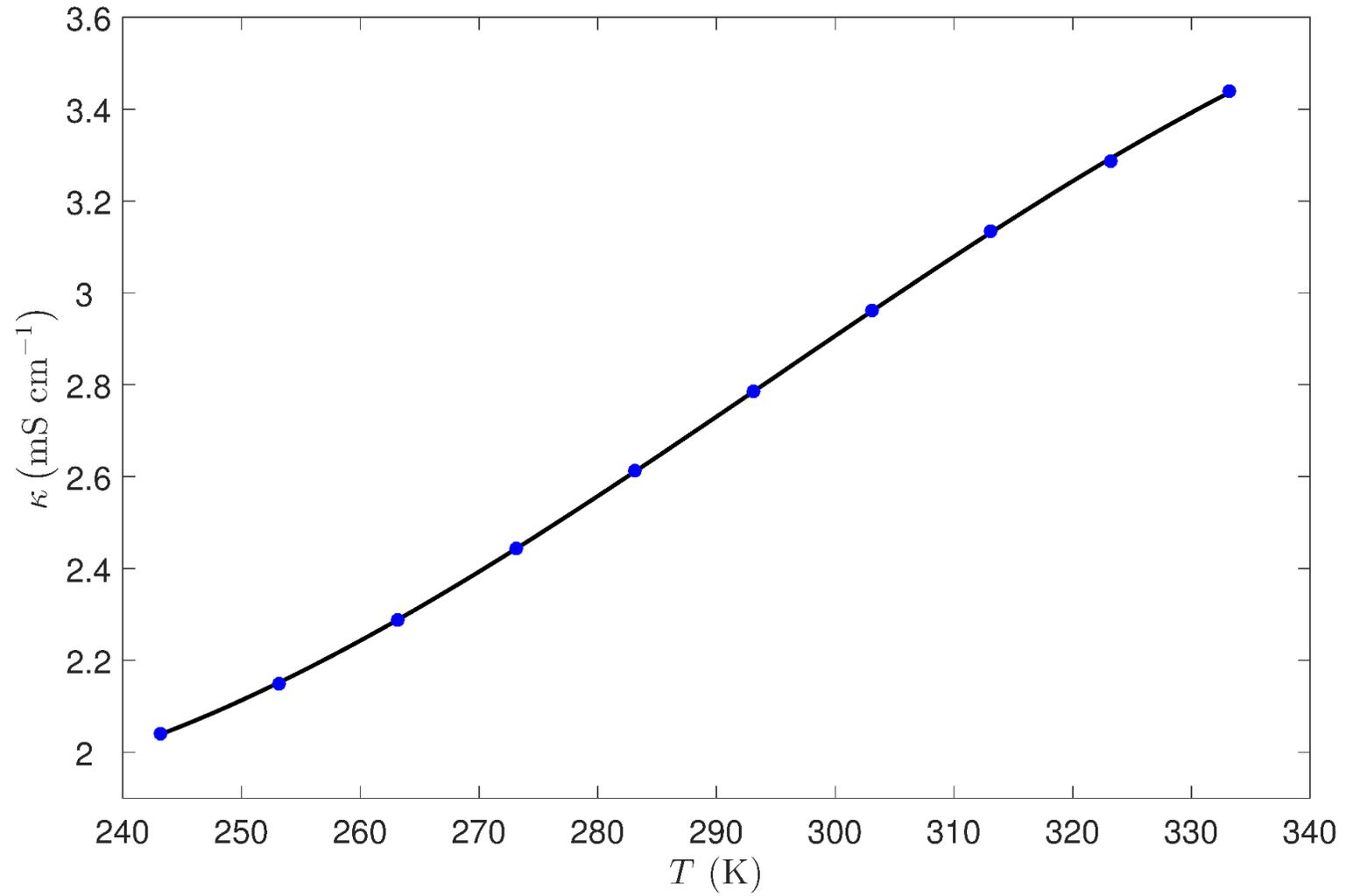

**Figure 4**

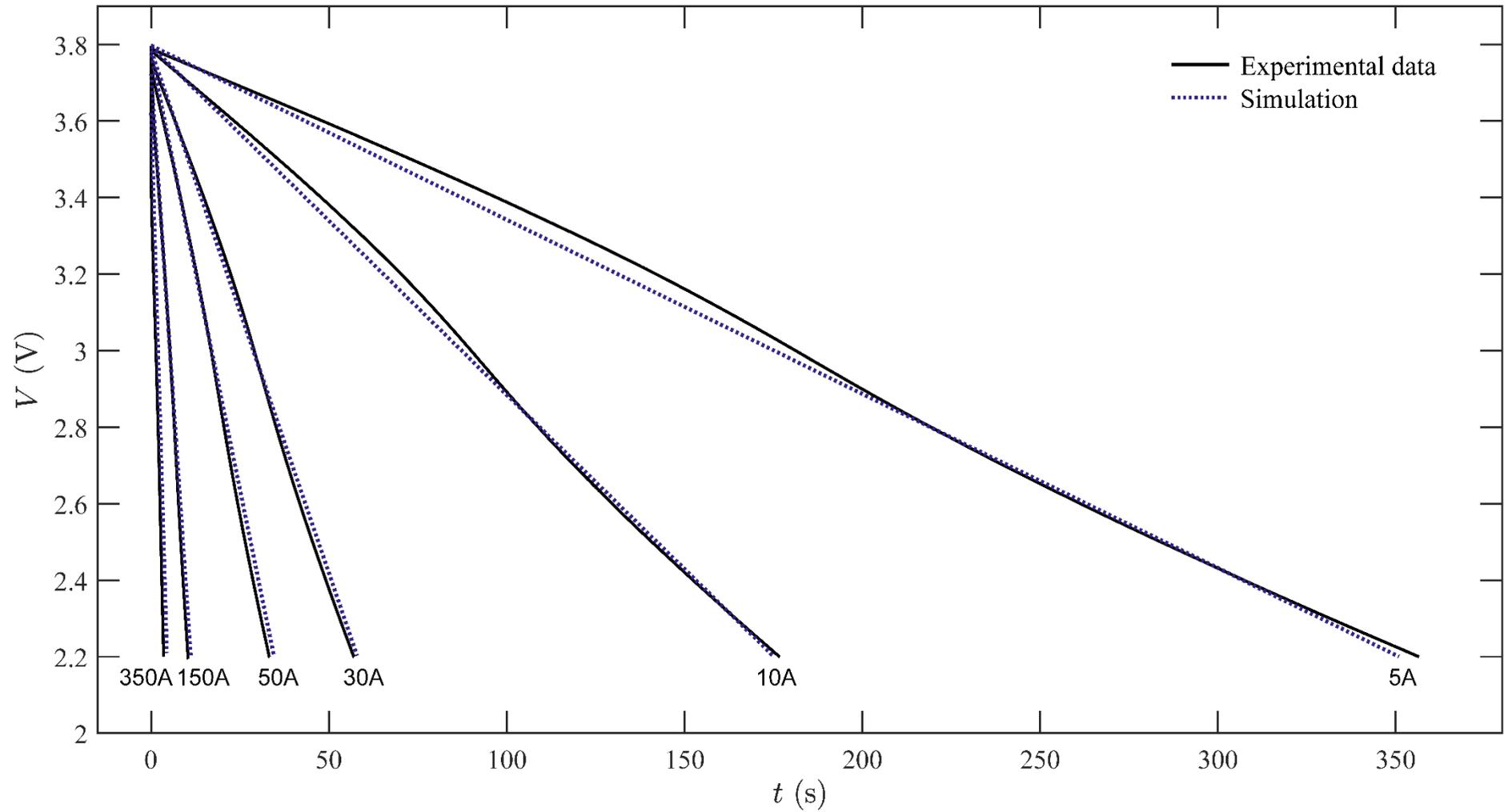

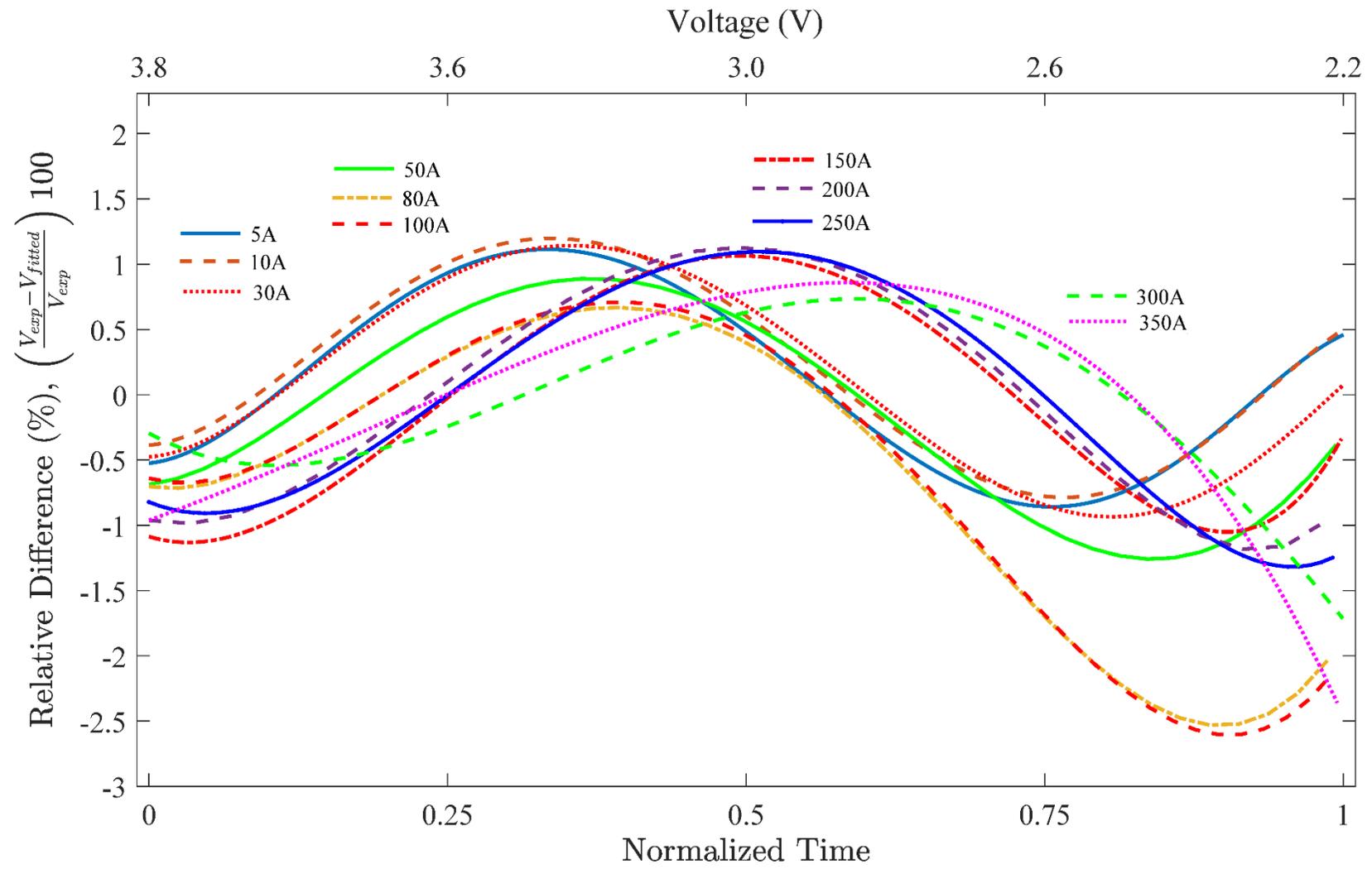

**Figure 6**

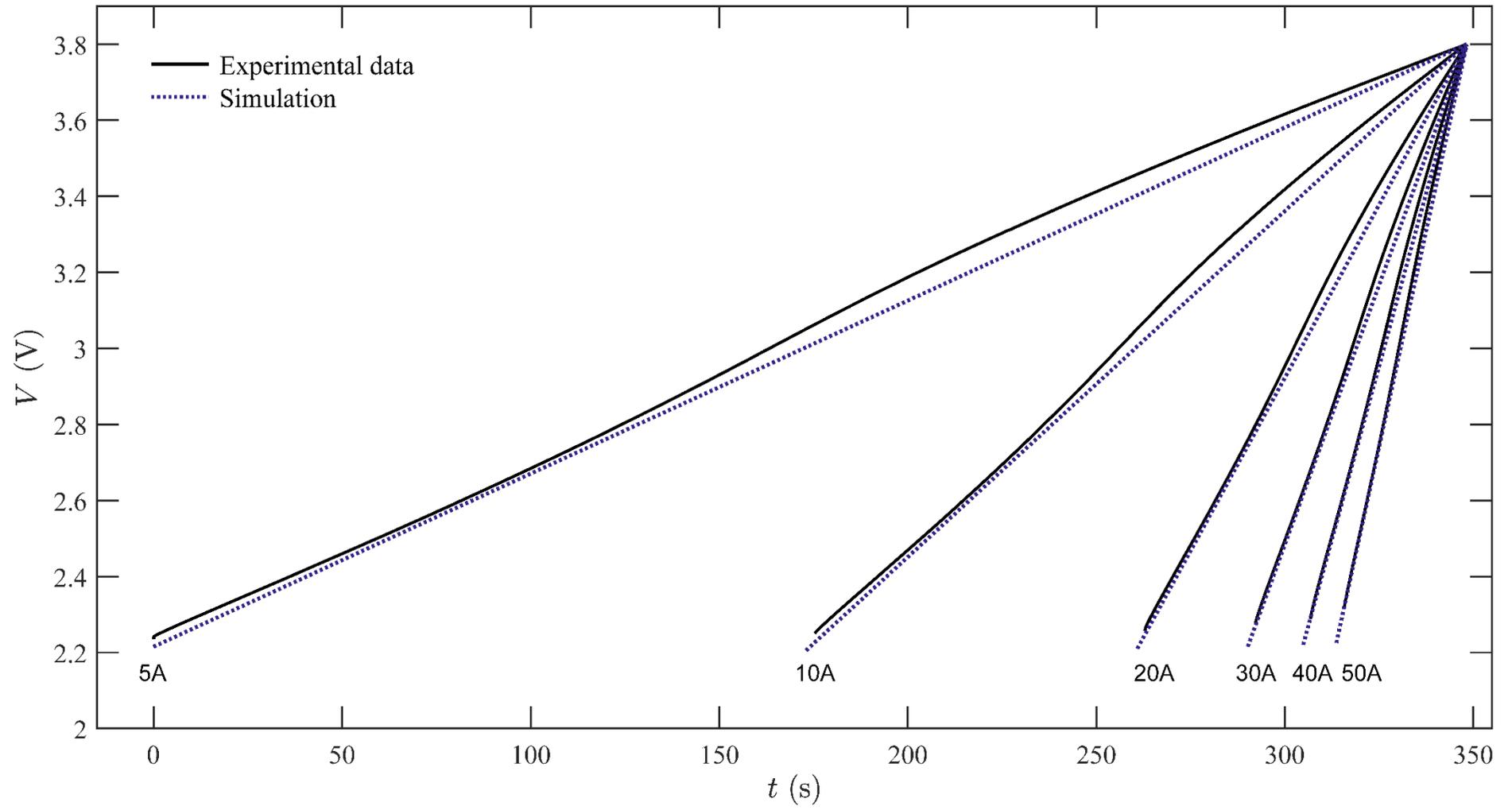

**Figure 7**

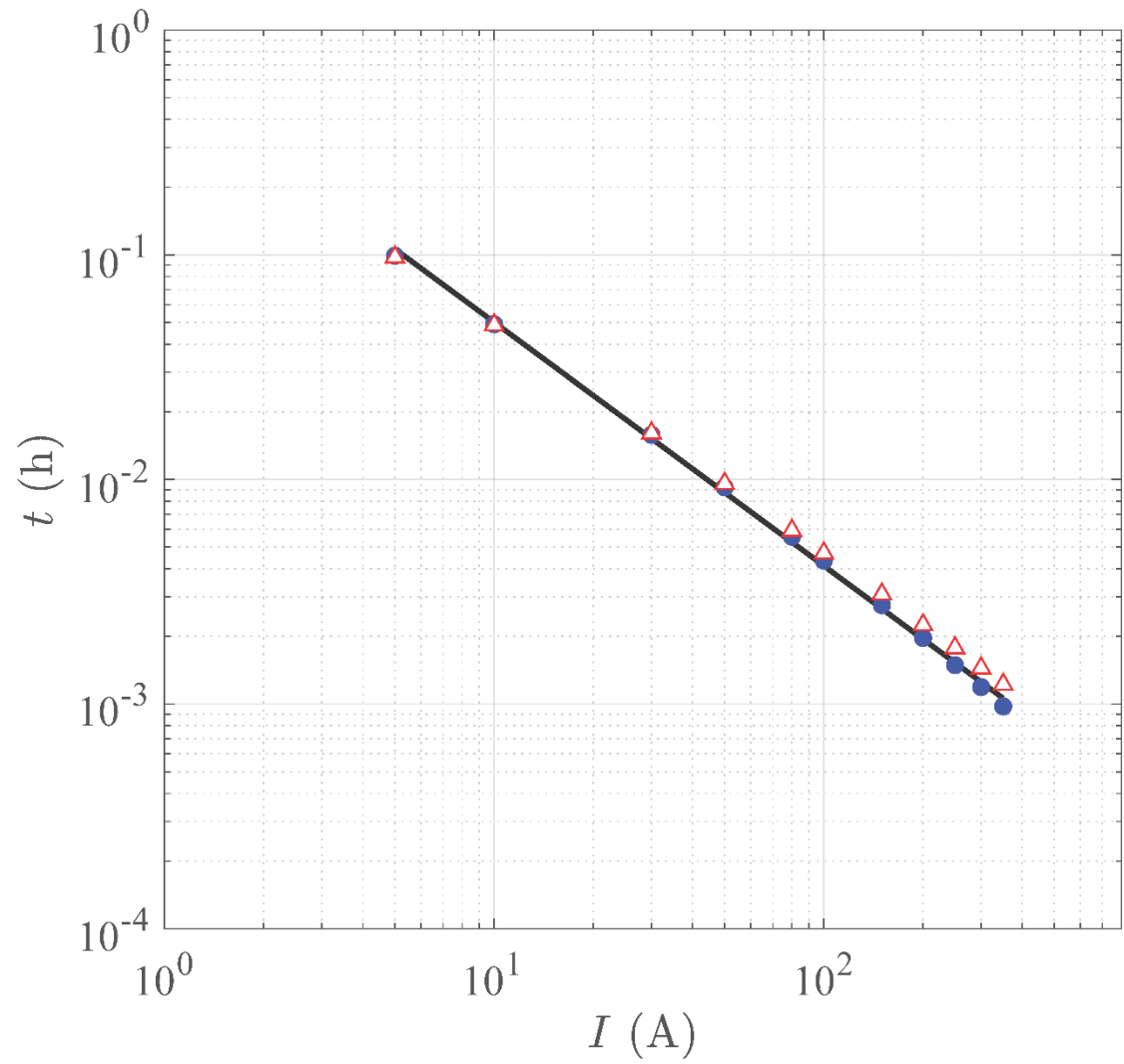

**Figure 8**

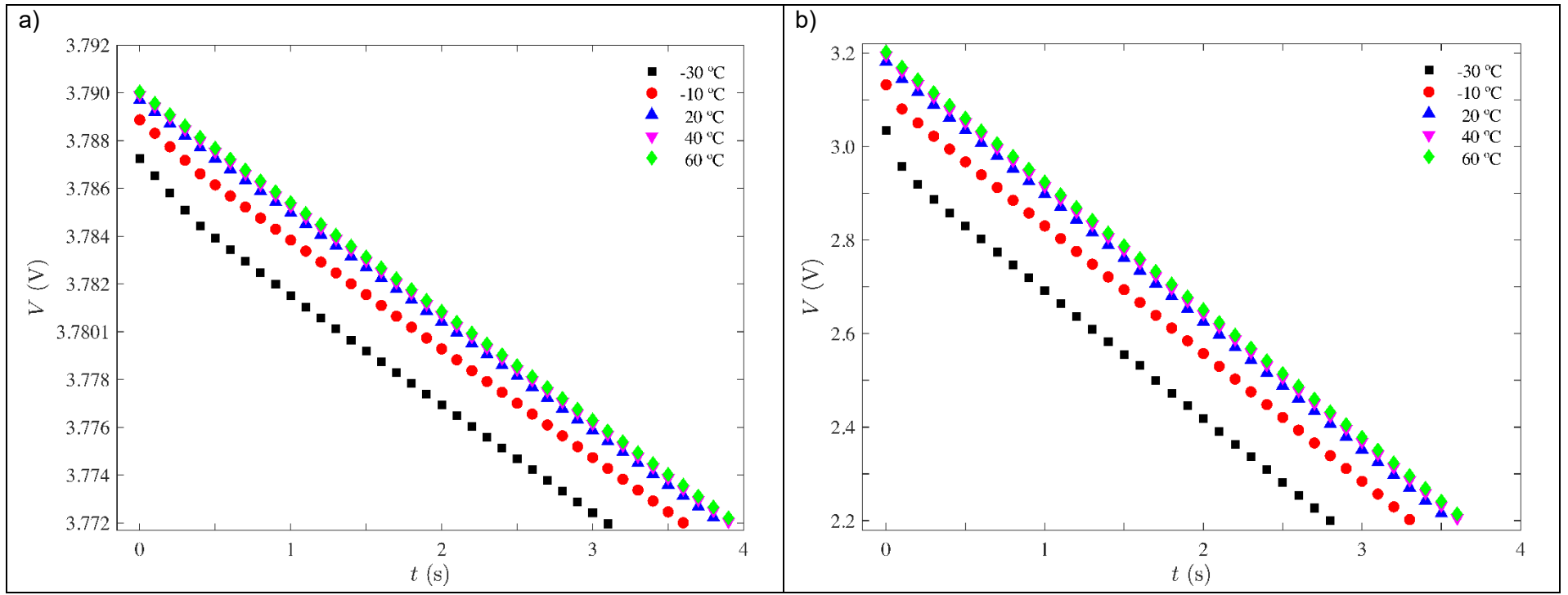

**Table captions**

**Table 1.** Specifications of the commercial LIC (JM Energy, Ultimo 1.100F).

**Table2.** List of geometrical, physical and chemical properties (298.15 K) of the inner elements of the LIC used to solve the electrochemical model.

**Table 3.** Experimental and theoretical electric data of galvanostatic discharging processes.

**Table 1**

| | |
|---|---|
| Range of operating temperatures | -30 °C / 70 °C |
| Maximum voltage | 3.8 V |
| Minimum voltage | 2.2 V |
| Maximum charge current | 50 A |
| Maximum discharge current | 360 A |
| Capacitance | 1100 F |
| Initial ESR (1 KHz) | 0.8 m$\Omega$ |
| Initial DC-IR (5 A discharge, 25°C) | 1.2 m$\Omega$ |
| Specific energy density | 10 Wh kg$^{-1}$ |
| Volumetric energy density | 19 Wh L$^{-1}$ |
| Weight | 145 g |
| Cell size | 180 $\times$ 126 $\times$ 5.5 mm |

| Name | Description | Value | Units | |
|---|---|---|---|---|
| $T$ | Temperature | 298.15 | K | |
| $L_-$ | Thickness of the negative electrode active material (graphite) | 33.0 | $\mu$m | measured |
| $L_+$ | Thickness of the positive electrode active material (activated carbon) | 50.5 | $\mu$m | measured |
| $L_{sep}$ | Thickness of the separator | 21.2 | $\mu$m | measured |
| $\sigma_0^-$ | Graphite bulk conductivity | $10^6$ | S m$^{-1}$ | [67] |
| $\sigma_0^+$ | Activated carbon bulk conductivity | 10 | S m$^{-1}$ | [67] |
| $\varepsilon_s^-$ | Solid phase volume-fraction negative electrode | 0.36 | | [68] |
| $\varepsilon_l^-$ | Electrolyte phase volume-fraction negative electrode | 0.52 | | [68] |
| $\varepsilon_s^+$ | Solid phase volume-fraction positive electrode | 0.33 | | [69] |
| $\varepsilon_l^+$ | Electrolyte phase volume-fraction positive electrode | 0.67 | | [69] |
| $\varepsilon_s^{sep}$ | Solid phase volume-fraction in separator | 0.7 | | [70, 71] |
| $U_{eq}$ | Equilibrium potential of faradaic electrode (graphite) | 0.1 | V | [72] |
| $R_{film}$ | Film resistance of the negative electrode | 0.1 | $\Omega$ m$^2$ | [73] |
| $i_0$ | Exchange current density | 25 | A m$^{-2}$ | [62, 74] |
| $\alpha_a$ | Anodic transfer coefficient | 0.5 | | [75] |
| $\alpha_c$ | Cathodic transfer coefficient | 0.5 | | [75] |
| $a_-$ | Volumetric specific surface area of graphite electrode | $2.26 \times 10^5$ | cm$^2$ cm$^{-3}$ | [67] |
| $a_+$ | Volumetric specific surface area of activated carbon electrode | $1.40 \times 10^7$ | cm$^2$ cm$^{-3}$ | [67] |
| $C$ | Double layer specific capacitance | $3.57 \times 10^{-2}$ | F m$^{-2}$ | [67] |

**Table 3**

| Applied current (A) | 5 | 10 | 30 | 50 | 80 | 100 | 150 | 200 | 250 | 300 | 350 |
|---|---|---|---|---|---|---|---|---|---|---|---|
| C-rate | 10C | 20C | 60C | 100C | 160C | 200C | 300C | 400C | 500C | 600C | 700C |
| Current density (mA cm$^{-2}$) | 1.15 | 2.30 | 6.89 | 11.49 | 18.38 | 22.98 | 34.47 | 45.96 | 57.45 | 68.93 | 80.42 |
| **Experimental** | | | | | | | | | | | |
| Discharge time (s) | 356.4 | 176.5 | 56.9 | 33.2 | 20.0 | 15.7 | 9.9 | 7.1 | 5.3 | 4.3 | 3.5 |
| Initial voltage (V) | 3.79 | 3.78 | 3.77 | 3.75 | 3.72 | 3.70 | 3.66 | 3.61 | 3.56 | 3.51 | 3.48 |
| Capacitance (F) | 1121 | 1113 | 1090 | 1073 | 1053 | 1042 | 1020 | 1003 | 983 | 974 | 963 |
| **Numerical analysis** | | | | | | | | | | | |
| Discharge time (s) | 350.9 | 175.1 | 58.0 | 34.6 | 21.4 | 17.0 | 11.1 | 8.2 | 6.4 | 5.3 | 4.4 |
| Initial voltage (V) | 3.80 | 3.79 | 3.78 | 3.77 | 3.76 | 3.75 | 3.72 | 3.70 | 3.67 | 3.65 | 3.62 |
| Capacitance (F) | 1099 | 1099 | 1098 | 1098 | 1097 | 1096 | 1094 | 1092 | 1092 | 1089 | 1089 |
| **Voltage** | | | | | | | | | | | |
| Maximum relative error (%) | 1.5 | 1.4 | 1.9 | 3.1 | 3.8 | 5.1 | 4.6 | 9.0 | 11.4 | 8.8 | 14.4 |
| Mean relative error (%) | 0.6 | 0.4 | 0.5 | 1.2 | 1.3 | 2.1 | 1.4 | 4.5 | 6.1 | 5.2 | 7.8 |